\documentclass[twocolumn,showpacs,preprintnumbers, prl,
               superscriptaddress]{revtex4}
\usepackage{graphicx, fancybox}
\usepackage{amsmath,amssymb}

\begin{document}


\title{
Third moments of conserved charges 
as probes of QCD phase structure
}

\author{Masayuki Asakawa}
\email{yuki@phys.sci.osaka-u.ac.jp}
\affiliation{
Department of Physics, Osaka University, Toyonaka, Osaka 560-0043, Japan}

\author{Shinji Ejiri}
\email{ejiri@quark.phy.bnl.gov}
\affiliation{
Physics Department, Brookhaven National Laboratory, Upton, NY 11973, USA}

\author{Masakiyo Kitazawa}
\email{kitazawa@phys.sci.osaka-u.ac.jp}
\affiliation{
Department of Physics, Osaka University, Toyonaka, Osaka 560-0043, Japan}

\begin{abstract}

The third moments of conserved charges, the baryon and electric charge 
numbers, and energy, as well as their mixed 
moments, carry more information on the state around the QCD phase
boundary than 
previously proposed fluctuation observables and higher order moments.
In particular, their signs give plenty of information on the
location of the state created in relativistic heavy ion collisions
in the temperature and baryon chemical potential plane.
We demonstrate this with an effective model.

\end{abstract}

\date{April 14, 2009}

\maketitle

Quantum chromodynamics (QCD) is believed to have a rich 
phase structure in the temperature ($T$) and baryon 
chemical potential ($\mu_{\rm B}$) plane.
Lattice QCD calculations indicate that 
the chiral and deconfinement phase transitions are a
smooth crossover on the temperature axis \cite{lat:pt}, 
while various models predict that the phase transition becomes 
of first order at high density \cite{Stephanov:2004wx}.
The existence of the QCD critical point is thus expected. 
To map these components of the phase diagram on the 
$T$-$\mu_{\rm B}$ plane is one of the most 
challenging and stimulating subjects which may be achieved 
by relativistic heavy ion collisions \cite{RHIC}.

Various observables have been proposed for this purpose 
\cite{Stephanov:1998dy,Asakawa:2000wh,Jeon:2000wg,Hatta:2003wn}.
Most scenarios suggested so far are concerned with
fluctuations, such as those of conserved charges, 
momentum distributions, slope parameters, 
and so forth.
For example, fluctuations of conserved charges
behave differently between the hadronic and quark-gluon plasma 
phases, and may be used as an indicator of the 
realization of the phase transition \cite{Asakawa:2000wh,Jeon:2000wg}.
The singularity at the critical point, at which the 
transition is of second order, may also cause enhancements 
of fluctuations if fireballs created by heavy ion 
collisions pass near the critical point during the time evolution
\cite{Stephanov:1998dy,Hatta:2003wn}.
Because of finite size effects and critical slowing down,
however, such singularities are blurred and its experimental 
confirmation
may not be possible \cite{Berdnikov:1999ph,Nonaka:2004pg}.
In fact, so far no clear evidence for the critical point has been
detected in event-by-event analyses \cite{RHIC}.
Approaches to use higher order moments for this purpose
have been also suggested recently \cite{higher} and
experimental attempts to measure those higher order
moments were reported, for example, in Ref.~\cite{Nayak:2009qm}.
Almost all previous studies, however, focus on the 
{\it absolute value}, especially the enhancement, 
of each observable around the phase boundary.

In the present Letter, we propose to employ {\it signs } of
third moments of conserved charges around the averages, 
which we call, for simplicity, the third moments in the following, 
to infer the states created by heavy ion collisions.
In particular, we consider third moments of conserved quantities,
the net baryon and electric charge numbers, and the energy,
\begin{align}
m_3( c c c )
\equiv \frac{ \langle ( \delta N_c )^3 \rangle }{ V T^2 }, 
\quad
m_3( {\rm EEE} )
\equiv \frac{ \langle ( \delta E )^3 \rangle }{ V T^5 },
\label{eq:ccc}
\end{align}
where $N_c$ with $c={\rm B,Q}$ represent 
the net baryon and electric charge numbers in a subvolume $V$, 
respectively, $E$ denotes the total energy in $V$, 
$\delta N_c= N_c - \langle N_c \rangle$, and 
$\delta E= E - \langle E \rangle$.
We also make use of the mixed moments defined as follows:
\begin{align}
m_3( cc{\rm E} )
\equiv \frac{ \langle ( \delta N_c )^2 \delta E \rangle }{ V T^3 },
\quad
m_3( c{\rm EE} )
\equiv \frac{ \langle \delta N_c ( \delta E )^2 \rangle }{ V T^4 }.
\label{eq:cee}
\end{align}

\begin{figure}[tbp]
\begin{center}
\includegraphics[width=.48\textwidth]{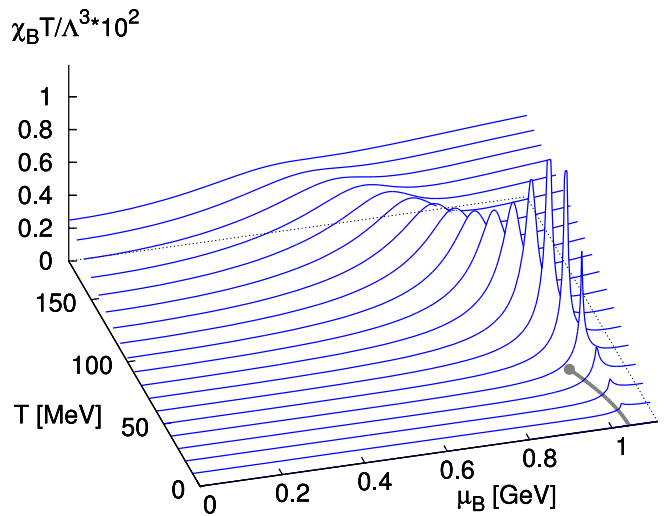}
\caption{
(color online).
$T$ and $\mu_{\rm B}$ dependence of the baryon number 
susceptibility $\chi_{\rm B}$ multiplied by $T$
in the Nambu-Jona-Lasinio model.
The bold line on the bottom surface shows the first order 
phase transition line and the point at the end is 
the critical point.
}
\label{fig:chi}
\end{center}
\end{figure}

To understand the behaviors of these moments around the QCD phase
boundary, we first notice that the moments Eqs.~(\ref{eq:ccc}) and 
(\ref{eq:cee}) are related to derivatives of the thermodynamic 
potential per unit volume, $\omega$, up to third order 
with respect to the corresponding chemical potentials and $T$.
The simplest example is $m_3({\rm BBB})$, which is given by
\begin{align}
m_3({\rm BBB})
= - \frac{\partial^3 \omega }{ \partial \mu_{\rm B}^3 }
= \frac{ \partial \chi_{\rm B} }{ \partial \mu_{\rm B} },
\label{eq:bbb}
\end{align}
where the baryon number susceptibility, $\chi_{\rm B}$, is defined as
\begin{align}
\chi_{\rm B} 
= - \frac{ \partial^2 \omega }{ \partial \mu_{\rm B}^2 }
= \frac{ \langle (\delta N_{\rm B} )^2 \rangle }{ VT }.
\end{align}
The baryon number susceptibility $\chi_{\rm B}$ diverges at the 
critical point and has a peak structure around there
\cite{Stephanov:1998dy,Hatta:2002sj,Fujii:2003bz}.
Since $m_3({\rm BBB})$ is given by the $\mu_{\rm B}$ derivative of
$\chi_{\rm B}$ as in Eq.~(\ref{eq:bbb}), the existence of the 
peak in $\chi_{\rm B}$ means that $m_3({\rm BBB})$ 
changes its sign there.
Although the precise size and shape of the critical region are not known, 
various models predict that the peak structure of $\chi_{\rm B}$ 
well survives 
far along the crossover line
\cite{Hatta:2002sj,Sasaki:2006ws,Stephanov:2004wx}
(See, Fig.~\ref{fig:chi} as a demonstration of this feature 
in a simple effective model; the details will be explained later).
This means that the near (hadron) and far (quark-gluon) sides of 
the QCD phase boundary can be distinguished by the sign of 
$m_3({\rm BBB})$ over a rather wide range around the critical point.
It is this feature that third moments carries more information
than fluctuations (second moments); fluctuations are, by definition,
positive definite and cannot differentiate the near side from the
far side as decisively as the third moments. 
We note that odd power moments around the averages in general
do not vanish except the first order one.
As we shall see later, 
all third moments presented in Eqs.~(\ref{eq:ccc}) and 
(\ref{eq:cee}) can be expressed in terms of derivatives of 
corresponding susceptibilities which diverge at the QCD critical 
point, and hence change their signs there.

The third moments can be measured in heavy ion 
collisions by the event-by-event analysis similarly 
to fluctuations, provided that $N_c$ and/or $E$ 
in a given rapidity range, $\Delta y$, in fireballs created by
collisions is determined in each event.
The measurement of $N_{\rm B}$ is difficult because of
the difficulty in identifying neutrons.
On the other hand,
$N_{\rm Q}$ and $E$ can be measured with the existing
experimental techniques.
Four out of the seven third moments in Eqs.~(\ref{eq:ccc}) and 
(\ref{eq:cee}) composed of $N_{\rm Q}$ and $E$ thus can
be determined experimentally.

All quantities we are considering here, $N_{\rm B,Q}$ and $E$, 
are conserved charges and the variation of their local densities
requires diffusion.
In Ref.~\cite{Asakawa:2000wh}, it was shown that the effect of
diffusion is small enough for the fluctuations of the baryon and
electric charges if the rapidity range is taken to be
$\Delta y \gtrsim 1$ \cite{Shuryak:2000pd}. 
In the estimate, the one dimensional
Bjorken expansion and straight particle trajectories were
assumed. If the contraction of hadron phase due to the
transverse expansion and the short mean free paths are
taken into account, the above estimate will be more relaxed.
This conclusion is not altered even if we take the effects of 
global charge conservation and resonance decays into account
\cite{Asakawa:2000wh,Jeon:2000wg,Bleicher:2000ek}.
On the other hand, the experimentally measured charge 
fluctuations at RHIC \cite{RHIC} are close to those in the 
hadron phase than the free quark-gluon one.
We, however, remark that the values of charge fluctuations 
in the quark-gluon phase can be similar to that in the 
hadron phase if the quark and gluons are strongly coupled 
\cite{Nonaka:2005vr}.
The experimental results therefore do not necessarily 
contradict a realization of the quark-gluon phase.

Once the negativeness of third moments is established experimentally, 
it is direct evidence of two facts:
(1) the existence of a peak structure of corresponding 
susceptibility in the phase diagram of QCD, and 
(2) the realization of hot matter beyond the peak, i.e. the
quark-gluon plasma, in heavy ion collisions.
We emphasize that this statement using the {\it signs} of third 
moments does not depend on any specific models.
The experimental measurements of signs of moments also have an 
advantage compared to their absolute values:
it is usually essential to normalize experimentally
obtained values by extensive observables, such as the total 
charged particle number $N_{\rm ch}$,
in order to compare the experimental results with theoretical 
predictions \cite{Asakawa:2000wh,Jeon:2000wg}.
In the measurement of signs, however,
normalization is not necessary.
In the measurement of absolute values, one has to be aware of 
the effect of global charge conservation
when $\Delta y$ is large \cite{Bleicher:2000ek}.
It is, however, expected that the effect does not change the 
signs of the third moments.
It is these features 
that our proposal is less 
subject to experimental and theoretical ambiguities
and more robust than previously proposed ones.

Let us now consider the behavior of third moments
other than $m_3({\rm BBB})$ around the critical point.
First, the third moment of the net electric charge 
$m_3({\rm QQQ})$ is calculated to be
\begin{align}
m_3({\rm QQQ})
&= - \frac{\partial^3 \omega }{ \partial \mu_{\rm Q}^3 }
\nonumber \\
&= -\frac18 \frac{ \partial^3 \omega }{ \partial \mu_{\rm B}^3}
- \frac38 \frac{ \partial^3 \omega }{ \partial \mu_{\rm B}^2 \mu_{\rm I} }
- \frac38 \frac{ \partial^3 \omega }{ \partial \mu_{\rm B} \mu_{\rm I}^2 }
- \frac18 \frac{ \partial^3 \omega }{ \partial \mu_{\rm I}^3}\, ,
\label{eq:QI}
\end{align}
where $\mu_{\rm Q}$ represents the chemical potential associated
with $N_{\rm Q}$, i.e. 
$\partial/\partial\mu_{\rm Q}
= (2/3)\partial/\partial\mu_{\rm u} 
- (1/3)\partial/\partial\mu_{\rm d}
= ( \partial/\partial\mu_{\rm B} + \partial/\partial\mu_{\rm I} )/2$,
and the isospin chemical potential is defined as 
$\mu_{\rm I} = ( \mu_{\rm u} - \mu_{\rm d} ) /2$ with 
$\mu_{\rm u,d}$ being the chemical potentials of the up and 
down quarks, respectively.
In relativistic heavy ion collisions, 
the effect of isospin symmetry breaking is small.
Assuming the isospin symmetry, the second and last terms in 
the most right hand side of Eq.~(\ref{eq:QI}) vanish and one 
obtains
\begin{align}
m_3({\rm QQQ})
&= \frac18 \frac\partial{ \partial \mu_{\rm B} }
\left( \chi_{\rm B} + 3 \chi_{\rm I} \right),
\label{eq:QI2}
\end{align}
with the isospin susceptibility 
$\chi_{\rm I} = -\partial^2 \omega / \partial \mu_{\rm I}^2$.
Under the isospin symmetry, $\chi_{\rm I}$ does not diverge 
at the critical point because the critical fluctuation does 
not couple to the isospin density \cite{Hatta:2003wn}.
The critical behavior of the term in the parenthesis in Eq.~(\ref{eq:QI2}) 
in the vicinity of the critical point is thus solely 
governed by $\chi_{\rm B}$.
Since $m_3({\rm QQQ})$ is a $\mu_{\rm B}$ derivative of
this term, a similar behavior as 
$m_3({\rm BBB})$ is expected.

Next, it can be shown that mixed moments including a single E 
are concisely given by
\begin{align}
m_3(cc{\rm E})
= \frac1T \left. \frac{ \partial (T \chi_c) }{ \partial T } 
\right|_{\hat\mu}, 
\label{eq:qqe}
\end{align}
with $c={\rm B}$, Q, where $\chi_{\rm Q} 
\equiv -\partial^2 \omega / \partial \mu_{\rm Q}^2
= ( \chi_{\rm B} + \chi_{\rm I} )/4 $ is the 
electric charge susceptibility.
The $T$ derivative in Eq.~(\ref{eq:qqe}) is taken 
along the radial direction from the origin with fixed
$\hat\mu\equiv\mu_{\rm B}/T$, i.e. $\partial/\partial T |_{\hat\mu}
= \partial/\partial T |_{\mu_{\rm B}}
+ (\mu_{\rm B}/T) \partial/\partial\mu_{\rm B} |_T$.
Since $T \chi_c$ diverges at the critical point,
Eq.~(\ref{eq:qqe}) again leads to a similar behavior of 
$m_3(cc{\rm E})$ as the above-mentioned moments.

To argue the behaviors of remaining third moments including 
two or three E's, it is convenient to first define 
$C_{\hat\mu} = -T ( \partial^2 \omega / \partial T^2 )_{\hat\mu} 
= \langle (\delta E)^2 \rangle / V T^2$.
The third moments are then given by
\begin{align}
m_3({\rm EEE}) &= \frac1{T^3} \left. 
\frac{ \partial ( T^2 C_{\hat\mu} ) }{ \partial T } 
\right|_{\hat\mu} ,
\label{eq:eee}
\\
m_3({\rm BEE}) &= 2 m_3({\rm QEE}) 
= \frac1T \frac{ \partial C_{\hat\mu}  }{\partial \mu_{\rm B} } \, .
\label{eq:qee}
\end{align}
Since $C_{\hat\mu}$ is the second derivative of $\omega$ 
along the radial direction, it diverges at the critical 
point which belongs to the same universality class as 
that of the 3D Ising model.
Therefore, $m_3({\rm EEE})$, $m_3({\rm BEE})$, and 
$m_3({\rm QEE})$, all change their signs at the critical point.

While the above arguments, based on the divergence of 
second derivative of $\omega$, guarantee the appearance
of the region with negative third moments in the vicinity 
of the critical point, they do not tell us anything about 
the size of these regions in the $T$-$\mu_{\rm B}$ plane.
In fact, all third moments considered here become positive
at sufficiently high $T$ and $\mu_{\rm B}>0$ where 
the system approaches a free quark and gluon system.
The regions are thus limited more or less near the 
critical point.

The information about the behavior of the third moments
at small $\mu_{\rm B}$ can be extracted from the 
numerical results in lattice QCD.
For example, with the Taylor expansion method 
the thermodynamic potential is calculated to be
$\omega = -c_2 (T) \mu_{\rm B}^2 -c_4 (T) \mu_{\rm B}^4
-c_6 (T) \mu_{\rm B}^6 - \cdots$, and one can read off the 
behavior of $m_3({\rm BBB})$ at small $\mu_{\rm B}$ as 
$m_3({\rm BBB}) = 24 [ c_4 (T) \mu_{\rm B} 
+ 5 c_6 (T) \mu_{\rm B}^3 + \cdots ] $.
Lattice simulations indicate that $c_4(T)$ is 
positive definite, while $c_6(T)$ becomes negative 
in the high temperature phase \cite{Allton:2005gk}.
From this result one sees that $m_3({\rm BBB})$ is 
positive for small $\mu_{\rm B}$, while 
the negative $c_6(T)$ suggests that the sign of 
$m_3({\rm BBB})$ eventually changes at sufficiently large 
$\mu_{\rm B}$.
Other moments for small $\mu_{\rm B}$ can also be 
evaluated in the Taylor expansion method by expanding 
$\omega$ with respect to $T$ and $\mu_{\rm Q}$.
If the contour lines of vanishing third moments
are close enough to the $T$-axis, the lattice simulations 
may be able to determine these lines. 
Since the region with a negative third moment should 
depend on the channel, combined information of signs of 
different third moments, and the comparison of the 
third moments obtained by experiments and lattice simulations, 
will provide a deep understanding about the state of the 
system in the early stage of relativistic heavy
ion collisions and the QCD phase diagram.

The range of $\mu_{\rm B}/T$ where lattice simulations are
successfully applied, however, is limited to small 
$\mu_{\rm B}/T$ with the present algorithms.
In particular, thermodynamics around the critical point
cannot be analyzed with the Taylor expansion method.
In order to evaluate the qualitative behavior of the third 
moments in such a region,
one has to resort to effective models of QCD.
To make such an estimate, here we employ the two-flavor 
Nambu-Jona-Lasinio model \cite{NJL,Kunihiro:1991qu} 
with the standard interaction 
${\cal L}_{\rm int} = G \{ (\bar\psi\psi)^2 
+ (\bar\psi i\gamma_5\tau_i \psi)^2 \}$,
where $\psi$ denotes the quark field.
For the model parameters, we take the values determined 
in Ref.~\cite{NJL}; $G=5.5$GeV$^{-2}$, the current quark 
mass $m=5.5$MeV, and the three-momentum cutoff $\Lambda=631$MeV.
For the isospin symmetric matter, this model gives a first 
order phase transition at large $\mu_{\rm B}$, as shown on
the bottom surface of Fig.~\ref{fig:chi} by the bold line.
The critical point is 
at $(T,\mu_{\rm B})\simeq(48,980)$MeV.

\begin{figure}[tbp]
\begin{center}
\includegraphics[width=.48\textwidth]{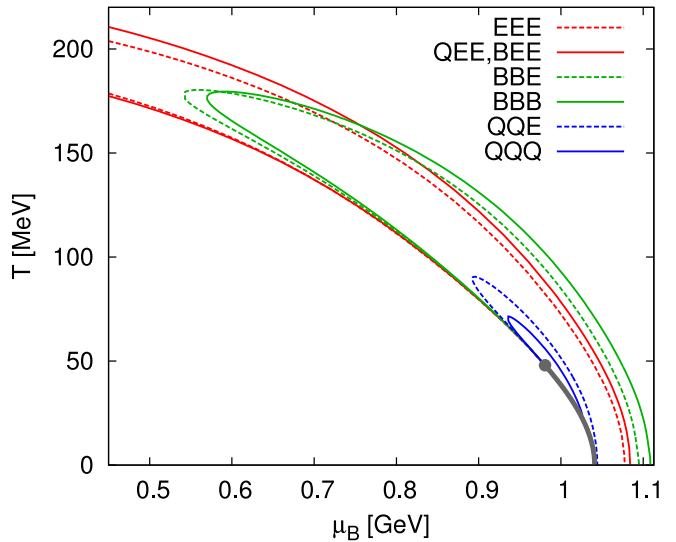}
\caption{
(color online).
Regions where third moments take negative values
in the $T$-$\mu_{\rm B}$ plane. The regions are inside the
boundaries given by the lines.
}
\label{fig:zero}
\end{center}
\end{figure}

In Fig.~\ref{fig:chi}, we also show the 
$T$ and $\mu_{\rm B}$ dependence of $T\chi_{\rm B}$
calculated in the mean-field approximation.
One observes that $\chi_{\rm B}$ diverges
at the critical point, and the peak structure well survives 
along the crossover line up to higher temperatures
\cite{Hatta:2002sj}.
The region where each moment becomes negative in the 
$T$-$\mu_{\rm B}$ plane is shown in Fig.~\ref{fig:zero}.
One sees that all
the moments become negative on the far side of 
the critical point as it should be, whereas the extent 
of the region depends on the channel.
The figure shows that areas with $m_3({\rm BBB})<0$ and 
$m_3({\rm BBE})<0$ extend to much lower $\mu_{\rm B}$ and
much higher $T$ than the critical point.
This suggests that even if the critical point is located at 
high $\mu_{\rm B}$ the negative third moments can be observed
by heavy ion collision experiments.
The figure also shows that the areas have considerable 
thicknesses along the radial direction.
Since the system stays near the phase transition line
considerably long regardless of the order of the phase transition,
first order or crossover, once the state on the far side is created,
negative third moments are very likely to be formed and observed.
The wide regions of negative moments also indicate that they
are hardly affected by critical slowing down 
and finite volume effects 
during the dynamical evolution of fireballs.

Figure~\ref{fig:zero} also shows that areas with negative 
$m_3({\rm EEE})$, $m_3({\rm QEE})$ and $m_3({\rm BEE})$ are 
much larger than those of the other moments
in the $T$-$\mu_{\rm B}$ plane;
although not shown in the figure, 
these areas extend even to the $T$-axis.
The behaviors of $m_3({\rm EEE})$ and $m_3(c{\rm EE})$ 
near the $T$-axis can be checked directly by the lattice 
simulations.
If the range of $T$ satisfying $m_3({\rm EEE})<0$ 
is sufficiently wide at $\mu_{\rm B}=0$, it is possible that
the negative third moments are measured even at the RHIC and 
LHC energies.
Whether the negative moments survive or not in this case
depends on the diffusion time of the energy density,
in other words the heat conductivity.
One can thus use the signs of $m_3({\rm EEE})$ and 
$m_3(c{\rm EE})$ to estimate the diffusion time of the 
charges and energy.
The third moments $m_3({\rm QQQ})$ and $m_3({\rm QQE})$,
on the other hand, become negative only in small regions 
near the critical point.
These behaviors come from the large contribution of 
$\chi_{\rm I}$ in Eq.~(\ref{eq:QI2}).

It should be, however, remembered that
the results in Figs.~\ref{fig:chi} and \ref{fig:zero},
are obtained in an effective model.
In particular, the model employed here gives 
the critical point at relatively low $T$ and high 
$\mu_{\rm B}$ \cite{Stephanov:2004wx}.
If the critical point is at much lower $\mu_{\rm B}$,
the areas with negative moments in Fig.~\ref{fig:zero}
should also move toward lower $\mu_{\rm B}$ and higher $T$.

In this Letter, we have pointed out that the third moments of
conserved charges, the net baryon and electric charge numbers and 
the energy,
carry more information on the state around the QCD phase
boundary than usual fluctuation observables.
They change signs at the phase boundary
corresponding to the existence of the peaks of susceptibilities.
If the negative third moments grow at early stage of the time 
evolution of fireball created in the collisions and 
if the diffusion of charges is slow enough, 
then the negative third moments will be measured experimentally
through event-by-event analyses.
Once such signals are measured, they serve direct evidence
that the peak structure of corresponding susceptibility exists 
in the phase diagram of QCD, and that the matter
on the far side of the phase transition, i.e. the quark-gluon
plasma, is created.
The combination of the third moments of different channels,
and their comparison with the
numerical results in lattice QCD
will bring various information on
the phase structure and initial states
created in heavy ion collisions at different energies.

This work is supported in part by a Grant-in-Aid for 
Scientific Research by Monbu-Kagakusyo of Japan 
(No.~19840037 and 20540268) and the U.S.\ Department of 
Energy (No.~DE-AC02-98CH10886).

\end{document}